\documentclass[aps,rmp,showpacs,twocolumn]{revtex4}
\usepackage{graphicx}
\begin{document}

%%%%%%%%%%%%%%%%%%%%%%%%%%%%%%%%%%%%%%%%%%%%%%%%%%%%%%%%%%%%%%%%%%%%%%%%%%%%%%%%%%%
\newcommand{\figureheight}{8.2 cm}
\newcommand{\putfig}[2]{\begin{figure}[h]
\special{isoscale #1.bmp, \the\hsize \figureheight}
\vspace{\figureheight} \caption{#2} \label{fig:#1}
\end{figure}}

% almost universal commands for equations and references
\newcommand{\eqn}[1]{(\ref{#1})}
\newcommand{\be}{\begin{equation}}
\newcommand{\ee}{\end{equation}}
\newcommand{\bea}{\begin{eqnarray}}
\newcommand{\eea}{\end{eqnarray}}
\newcommand{\bean}{\begin{eqnarray*}}
\newcommand{\eean}{\end{eqnarray*}}
\newcommand{\nn}{\nonumber}
%%%%%%%%%%%%%%%%%%%%%%%%%%%%%%%%%%%%%%%%%%%%%%%%%%%%%%%%%%%%%%%%%%%%%%%%%%%%%%%%%%%
%\draft

\title{ Transport through a double barrier in Large Radius  Carbon Nanotubes in
the presence of a transverse magnetic field }
\author{S. Bellucci $^1$ and P. Onorato $^1$ $^2$ \\}
\address{
$^1$INFN, Laboratori Nazionali di Frascati,
P.O. Box 13, 00044 Frascati, Italy. \\
$^2$Dipartimento di Scienze Fisiche, Universit\`{a} di Roma Tre,
Via della Vasca Navale 84, 00146 Roma, Italy}
\date{\today}
%\maketitle %\widetext
\pacs{05.60.Gg, 71.10.Pm, 73.63.-b, 71.20.Tx, 72.80.Rj}
\begin{abstract}
We discuss the Luttinger Liquid behaviour of  Large Radius Carbon
Nanotube e.g. the  Multi Wall ones (MWNT), under the action of a
transverse magnetic field $B$. Our results imply a reduction with
$B$ in the value of the $bulk$ critical  exponent,
$\alpha_{bulk}$, for the tunneling density of states, which is in
 agreement with that observed in transport experiments.

Then, the problem of the transport  through a Quantum Dot formed
by  two intramolecular tunneling barriers along the MWNT, weakly
coupled to Tomonaga-Luttinger liquids is studied, including the
action of a strong transverse magnetic field $B$. {We predict the
presence of some peaks in the conductance G versus $B$, related to
the magnetic flux quantization in the ballistic regime (at a very low
temperature, $T$) and also  at higher $T$, where the Luttinger
behaviour dominates}. The temperature dependence of the maximum
$G_{max}$ of the conductance peak according to the Sequential
Tunneling follows a power law, $G\propto T^{\gamma_e-1}$ with
$\gamma_e$ linearly dependent on the critical exponent,
$\alpha_{end}$, strongly reduced by $B$.

\end{abstract}

\maketitle

\section{Introduction}

In a recent paper\cite{noiqw} we discussed the transport through a
double barrier for interacting quasi one-dimensional electrons in
a Quantum Wire (QW), in  the presence of a transverse magnetic
field. Here we want to extend the results obtained there to an
analogous device based on Large Radius Carbon Nanotubes (LRCN), such as
the Multi Wall ones (MWNT). This aim is not trivial to pursue,
because of the geometry-dependent electronic properties of
Carbon Nanotubes (CNs) and the effects of many subbands
crossing the Fermi level in LRCNs.

\

{\it Transport in 1 Dimension} - Electronic correlations have been
predicted to dominate the characteristic features in quasi one
dimensional (1D) interacting electron systems. This  property,
commonly referred to as
 Tomonaga-Luttinger liquid (TLL)
behaviour\cite{TL}, has  recently moved into the focus of
attention by physicists, also  because in recent years several
electrical transport experiments for a variety of 1D devices, such
as semiconductor quantum wires\cite{wires} (QWs) and carbon
nanotubes (CNs)\cite{cnts} have shown this behaviour.

%%%%%%%%%%%%%%%%%%%%%   interazione 1D e LL
In a 1D electron liquid  Landau quasiparticles are unstable % in 1D systems of
%interacting electrons, in which
and the low-energy excitations take the form of plasmons
(collective electron-hole pair modes)%that are the stable
%excitations
: this is known as the  breakdown of the Fermi liquid picture in 1D. The LL
state has two main features: i) the power-law dependence of
physical quantities, such as the tunneling density of states
(TDOS), as a function of energy or temperature; ii) the
spin-charge separation: an additional electron in the LL decays
into decoupled spin and charge wave packets, with different
velocities for charge and spin. It follows that 1D electron
liquids are characterized by  the power-law dependence of some
physical quantities as a function of the energy or the
temperature. Thus the tunneling conductance $G$ reflects the power
law dependence of the DOS in  a small bias experiment\cite{kf}
 \bea
 G=dI/dV\propto
T^{\alpha_{bulk}} \eea  for $eV_b\ll k_BT$, where $V_b$ is the
bias voltage, $T$ is the temperature and $k_B$ is Boltzmann's
constant.

\

The power-law behaviour characterizes also the thermal dependence
of $G$ when an impurity is present along the 1D devices. The
theoretical approach to the presence of obstacles  mixes two
theories corresponding to  the single particle scattering (by a
potential barrier $V_B({\bf r})$) and the TLL theory of
interacting electrons. The single particle scattering gives the
transmission, probability, $|t|^2$, depending in general on  the
single particle energy $\varepsilon$. Hence, following
ref.\cite{sh},  the conductance, $G$, as a function of  the
temperature and $|t|$ can be obtained
\begin{equation} \label{gsp}
G\propto|t(\varepsilon, T)|^2\equiv |t(\varepsilon)|^2 T^{2
\alpha_{end},}
\end{equation}
where we introduced a second critical exponent, $\alpha_{end}$.

 \

{\it Intrinsic Quantum Dot -} Experiments\cite{Postma01,Bozovic01}
show transport through an intrinsic quantum dot (QD) formed by a
double barrier within a 1D electron system, allowing for the study
of the resonant or sequential tunneling. The linear conductance
typically displays a sequence of peaks, when the gate voltage,
$V_g$, increases. Thus also the double-barrier problem  has
attracted a significant amount of attention among
theorists\cite{Sassetti95,Furusaki98,Braggio00,Thorwart02,Nazarov03,Polyakov03,Komnik03,Hugle04},
in particular for the case of  two identical, weakly scattering
barriers at a distance $d$. In general, the transmission is
non-zero  for particular values of the parameters corresponding to
a momentum $k_F$, such that $\cos(k_F d/2)=0$.{ It follows that,
although in a 1D electron system  for repulsive interaction the
conductance is suppressed at zero temperature by the presence of
one impurity (1D metal becomes a perfect insulator), the presence
of an intrinsic QD  gives rise to some peaks in the conductance at
$T=0$ corresponding to the perfect transmission. This {\em
resonant scattering} condition corresponds to an average particle
number between the two barriers of the form $\nu+1/2$, with
integer $\nu$, i.e. the ``island'' between the two barriers is in
a degenerate state. If interactions between the electrons in the
island are included, one can recover the physics of the Coulomb
blockade\cite{kf,furusaki_double_barriere}.

The power-law behaviour characterizes also the thermal dependence
of $G$ in the presence of an IQD. A first theory about the
transport through an IQD is known as {\it Uncorrelated Sequential
Tunneling} (UST), where an incoherent sequential tunneling is
predicted. It follows the dependence of the peaks of the
conductance according to the power law
$$
 G_{max}\propto T^{\alpha_{end}-1}.
$$

%intro 2 barriere
Some  experiments\cite{Postma01,Bozovic01} showed transport
through an intrinsic quantum dot (QD) formed by a double barrier
within a Single Wall CN (SWNT), allowing one to study the resonant
or sequential tunneling. In order to explain
 the unconventional power-law
dependencies in the measured transport properties of a CN, a
mechanism was proposed\cite{Postma01,Thorwart02}, namely,
{\em correlated sequential tunneling} (CST) through the island.
The temperature dependence of the maximum $G_{max}$ of the
conductance peak, according to the CST  theory, yields the power
law behaviour \bea \label{gt2}
 G_{max}\propto T^{\alpha_{end-end}-1}=T^{2\alpha_{end}-1}.
\eea

Recently a lot of theoretical work has been carried out on the
double impurity problem in TLL systems. In an intermediate
temperature range $\varepsilon_c  \ll k_B T \ll \Delta_{dot}$,
where   $\varepsilon_c$ is the Infra Red cut-off energy and
$\Delta_{dot}$ is the level spacing of the dot, some
authors\cite{Nazarov03,Polyakov03} predict a behaviour according
to the UST, while others\cite{Hugle04} find results in agreement
with the CST theory. In a recent paper\cite{meden} the authors
discussed how the critical exponent can depend on the size of the
dot and on the temperature, by identifying three different
regimes, i.e. the UST at low $T$, a Kirchoff regime at
intermediate $T$ ($G_{max}\propto T^{2 \alpha_{end}}$) and a third
regime for $ T \gg \Delta_{dot}$, with $G_{max}\propto T^{-1}$. Thus,
in their calculations, obtained starting from spinless fermions on
the lattice model, no evidence of CST is present.

\

{\it Multi Wall Carbon Nanotubes -} An ideal Single Wall CN (SWCN)
is an hexagonal network of carbon atoms (graphene sheet) that has
been rolled up, in order to make a cylinder with a radius about $1
nm$ and a length about $1 \mu m$. The unique electronic properties
of CNs are due to their diameter and chiral angle
(helicity)\cite{3n}. MWCNs, instead, are made by several
(typically 10) concentrically arranged graphene sheets with a
radius above $5 nm$ and a length which ranges from  $1$ to some
hundreds of $\mu m$s. The transport measurements carried out in
the MWNTs reflect usually the electronic properties of the outer
layer, to which the electrodes are attached. Thus, in what follows
we mainly discuss the LRCNs as a general class of CNs including
also MWNTs.
 In
general  the LRCNs are  affected by the presence of doping,
impurities, or disorder, what leads to the presence of a large
number of subbands, $N$, at the Fermi level\cite{[21]}. It follows
that the critical exponent has a different form with respect to that
calculated in ref.\cite{noiqw}.

The bulk critical exponent can be calculated in several
different ways, e.g. see ref.\cite{npb} where  we obtained
\begin{equation}\label{al1n}
  \alpha_{bulk}\approx \frac{1}{4 N} \left(K_N+\frac{1}{K_N}-2 \right),
\end{equation}
where
$$
\frac{1}{K_N}\approx \sqrt{1 + \frac{N U_0(q_c,B)}{ (2 \pi
{v}_F)}}.
$$
Here $v_F$ is the Fermi velocity, $ U_0(p) $ corresponds to the
Fourier transform of the 1D e-e interaction potential  and
$q_c=2\pi/L$ is the infra-red natural cut-off due to the length of
the CN, $L$. For a strictly 1D system, such as a CN in absence of
magnetic field, $U_0(p)$ does not depend on the momenta of the
interacting electrons. In general\cite{noiqw0} we need to
introduce two different couplings for two different forward
scattering  processes (with a small transferred momentum). The
first term, $g_2$, is obtained by considering  $2$ scattered
electrons with opposite momenta ($\pm k_F$). The second term,
$g_4$, is obtained by considering  $2$ scattered electrons with
(almost) equal momenta ($k_1\sim k_2 \sim k_F$). It follows that
$$
{K_N}\approx \sqrt{\frac{2 \pi {v}_F+N \left(g_4-g_2\right)/2}{2
\pi {v}_F+N \left(g_4+g_2\right)/2 }},
$$
which  corresponds to the previous formula when
$g_2=g_4=U_0(q_c)$.  As in ref.\cite{noiqw0} the presence of a
magnetic field gives $g_2\neq g_4$, because of the edge
localization of the currents with opposite chiralities, and we need
the $B$ dependent values of $g_2$ and $g_4$.

The value of $\alpha_{bulk}$ obtained in ref\cite{npb}  is in
agreement with the one obtained in ref.\cite{egmw} where also the
end critical exponent was obtained as
\begin{equation}\label{al2n}
  \alpha_{end}\approx \frac{1}{2 N} \left(\frac{1}{K_N}-1 \right).
\end{equation}
} \

{\it Power law in MWNTs -}One of the most significant observations
made in the MWNTs has been the power-law behavior of the tunneling
conductance as a function of the temperature or the bias voltage.
The measurements carried out in the MWNTs have displayed a
power-law behavior of the tunneling conductance, that gives a
measure of the low-energy density of states. Although the power
law behaviour in the temperature dependence of $G$ usually
characterizes a small range of temperature (from some $-K$ up to
some tens, rarely up to the room $T$), this behaviour allows for the
measurement of the  critical  exponent $\alpha_{Bulk}$ ranging,
in MWNTs, from $0.24$ to $0.37$\cite{B}. These values are, on the
average, below those measured in single-walled nanotubes,
which are typically about $0.35$ \cite{y}. A similar behaviour
was satisfactory explained\cite{npb} in terms of the number of
subbands by applying eq.(\ref{al1n}).

 \

%%%%%%%%%%%%%%%%%%%%% QW w NT

{\it CNs under a transverse magnetic field} - The effects of a
transverse magnetic field $B$, acting on CNs were also
investigated in the last years. Theoretically, it is predicted
that a perpendicular $B$ field modifies the DOS of a CN
\cite{[33]}, leading to the Landau level formation. This effect
was observed in a MWNT single-electron transistor \cite{[34]}. In
a recent letter Kanda et al.\cite{kanda} examined the dependence
of $G$ on perpendicular $B$ fields in MWNTs. They found that, in
most cases, $G$ is smaller for higher magnetic fields, while
$\alpha_{Bulk}$ is reduced  by a factor $1/3$ to $1/10$, for $B$
ranging from $0$ to $4$ T. Recently we discussed the effects of a
transverse magnetic field in QWs\cite{noiqw} and large radius
CNs\cite{noimf}. The presence of $B\neq0$ produces the rescaling
of all repulsive terms of the interaction between electrons, with
a strong reduction of the backward scattering, due to the edge
localization of the electrons. Our results imply a variation with
$B$ in the value of $\alpha_{Bulk}$, which is in fair agreement
with the value observed in transport experiments\cite{kanda}.

\

{\it Impurities, buckles  and Intrinsic QD - }The magnetic induced
localization of the electrons should have some interesting effects
also on  the backward scattering, due to the presence of one or
more  obstacles along the LRCN, and hence on the corresponding
conductance, $G$\cite{noiqw}. Thus, the main  focus of our paper is
to analyze the presence of  two barriers along a LRCN     at a
fixed distance $d$. A similar device was made by the manipulation
of individual nanotubes with an atomic force microscope which
permitted the creation of intratube buckles acting as tunneling
barriers\cite{Postma01}. The SWNTs with two intramolecular buckles
have been reported to behave as a room-temperature single electron
transistor. The linear conductance typically displays a sequence
of peaks when the gate voltage, $V_g$, increases. The
one-dimensional nature of the correlated electrons is responsible
for the differences to the usual quantum Coulomb blockade theory.

We predict that, in the presence of a transverse magnetic field,
 a LRCN should show some oscillations
in the conductance as a function of the magnetic field, like those
discussed in ref.\cite{noiqw}.

%%%%%%%%%%%%%%%%%%%%%%%%%%%%%%%   SOMMARIO   %%%%%%%%%%%%%%%%%%%%%%%%%%%%%%%%%%%%%%

\

{\it Summary -} In this paper we want to discuss the  issues
mentioned above. In order to do that we follow the same structure
of our previous paper\cite{noiqw}.

In section II we introduce a theoretical model which can describe
the CN under the effect of a transverse magnetic field, and we
discuss the properties of the interaction starting from the
unscreened long range Coulomb interaction in two dimensions.

In section III we evaluate the $bulk$ and $end$ critical exponents.
Then  we discuss the effects on them due to an increasing
transverse magnetic field. We remark that $\alpha_{bulk}$
characterizes the discussed power-law behavior of the TDOS, while
($\alpha_{end}$)  characterizes the temperature dependence of
$G_{max}$, in both the UST and the CST regime.
  Finally, we discuss the presence  of an
intrinsic QD and the magnetic field dependent oscillations in the
conductance.

\section{Model and Interaction}

{\it Single particle - } Starting from the  known bandstructure of
graphite, after the definition of the boundary condition (i.e. the
wrapping vector $\overrightarrow{w}=(m_w,n_w)$), it is easy to
calculate the bandstructure of a CN. For an armchair CN
($m_w=n_w$) we obtain that the energy vanishes for two different
values of the longitudinal momentum $ \varepsilon_0(\pm K_s)=0$.
After fixing the angular momentum along the $y$ direction to be
$m\hbar$, the dispersion law $\varepsilon_0({m}, k)$ is usually
taken to behave linearly, so that we can approximate it as
$\varepsilon_0({m}, k)\approx\varepsilon_0({m},K_s)+v_F|k-K_s|$,
%$$
%\varepsilon_0({m}, k) \approx \pm \gamma\left(\frac{\pi
%m}{N_b}+\frac{3}{2}\left[|k-K_0|-\frac{|k-K_0|^2}{4}\right]\right)
%$$
where we  introduce a Fermi velocity $v_F%=3a \gamma/(2\hbar)
$ (about $10^6\, m/s$ for CNs). In general, we can define an
approximate one-dimensional bandstructure for momenta near $\pm
K_s=\pm (2\pi)/(3 a_0)$ \bea \varepsilon_0({m},\overrightarrow{w},
k)\approx \pm \frac{ v_F \hbar}{R}
\sqrt{\left(\frac{m_w-n_w+3m}{3}\right)^2 +R^2\left(k \pm
K_s\right)^2}\label{bs0} \eea where $R \approx N_b \sqrt{3}
a/(2\pi)$ is the tube radius (about $5 \, nm$ for MWNTs) and $a$
denotes the honeycomb lattice constant ($a/\sqrt{3} =a_0=
1.42\AA$).

For a metallic CN (e.g. the armchair one with $m_w=n_w$) we obtain
that the energy vanishes for two different values of the
longitudinal momentum $ \varepsilon_0(\pm K_s)=0$. The dispersion
law $\varepsilon_0({m}, k)$ in the case of undoped metallic
nanotubes is quite linear near the crossing values $\pm K_s$. The
fact of having four low-energy linear branches at the Fermi level
introduces a number of different scattering channels, depending on
the location of the electron modes near the Fermi points.

Starting from eq.(\ref{bs0}) we can develop a Dirac-like theory
for CNs corresponding to the hamiltonian
  \bea\label{hd}
H_D=v_F\left[\widehat{\alpha}(\hat{L}_z)+\widehat{\beta}\hat{\pi}_y\right],
\eea with a solution in the spinorial form $\widehat{\psi}$ where
\bea \widehat{\alpha}=\alpha\left(\begin{array}{cc}
  0 & i \\
  -i & 0
\end{array}\right) \;\;\; \widehat{\beta}=\left(\begin{array}{cc}
  0 & 1 \\
  1 & 0
\end{array}\right) \;\;\; \widehat{\Psi}=\left(\begin{array}{c}
  \psi_\uparrow  \\
  \psi_\downarrow
\end{array}\right). \eea
Here $\hat{\pi}_y=\hat{p}_y\pm\hbar K_s$, and
$\alpha=\frac{1}{R^2}$, and eq.(\ref{hd}) can be compared with the
one obtained in ref\cite{susy}.

For the metallic CN, such as the armchair one, the problem in eq.(\ref{hd})
has periodic boundary conditions {\em i.e. }
$\Psi(\varphi+2\pi,y)=\Psi(\varphi)$, it follows that a factor
$e^{i m \varphi}$ appears in the wavefunction. For semiconducting
CNs ($m_w\neq n_w$) we have to define {\it quasiperiodic  boundary
conditions  i.e. } $\Psi(\varphi+2\pi,y)=\omega
\Psi(\varphi)$\cite{susy} corresponding to a factor $e^{i
(m+\frac{m_w -n_w}{3} \varphi)}$  in the wavefunction ($m_0=m_w
-n_w$).

A cylindrical carbon nanotube {with the axis along the $y$
direction and} $B$ along $z$ corresponds to \bea\label{hd1}
H_D=v_F\left[\widehat{\alpha}(\hat{L}_z)+\widehat{\beta}\left(\hat{\pi}_y-\frac{e}{c}{\bf
A }\right)\right], \eea where we choose the gauge so that the
system has a symmetry along the $\hat{y}$ direction, $${\bf
A}=(0,Bx,0)=(0,B R \cos(\varphi),0)$$ and  we introduce the
cyclotron frequency $\omega_c=\frac{eB}{m_e c}$  and the magnetic
length $\ell_\omega=\sqrt{\hbar/(m \omega_c)}$.

It is usual to discuss the results in terms of two parameters, one
for the scale of the energy following from eq.(\ref{bs0}) \be
\label{def-delta} \Delta_0={\hbar v_F \over R} \ , \ee the second one
being the scale of the magnetic field \be \label{def-A} \nu\equiv
{\pi R^2 \over 2\pi \ell_\omega^2}= {\pi R^2B \over \Phi_0} \quad
where \quad  \Phi_0={hc \over e}. \ee

Here we can calculate the effects of the magnetic field by
diagonalizing eq.(\ref{hd1}), after introducing the  trial
functions
 \bea\label{ft}
 \widetilde{\psi}_{s,m, k}(\varphi,y)=N e^{i\left(k y+(m+m_0)
\varphi\right)}\left(\alpha_s+ \beta_s \sin(\varphi)+\gamma_s
\cos(\varphi)) \right). \eea Results are reported in Fig.(1) for
different CNs and values of the magnetic field.
%%%%%%%%%%%%%%%%%%%%%%%%%%%%%%%%%%%%%%%%%%%%%%%%%%%%%%%%%%%%%
\begin{figure}
\includegraphics*[width=0.70\linewidth]{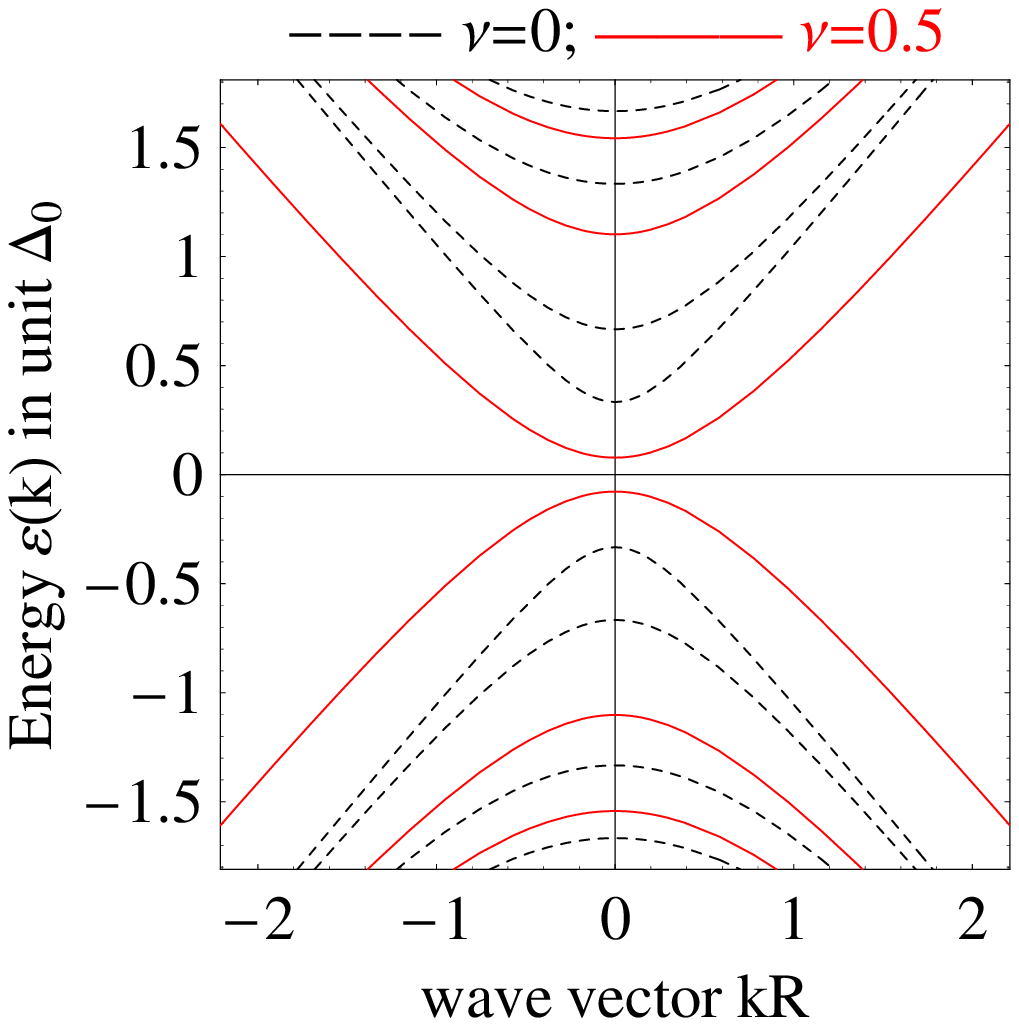}
\includegraphics*[width=0.70\linewidth]{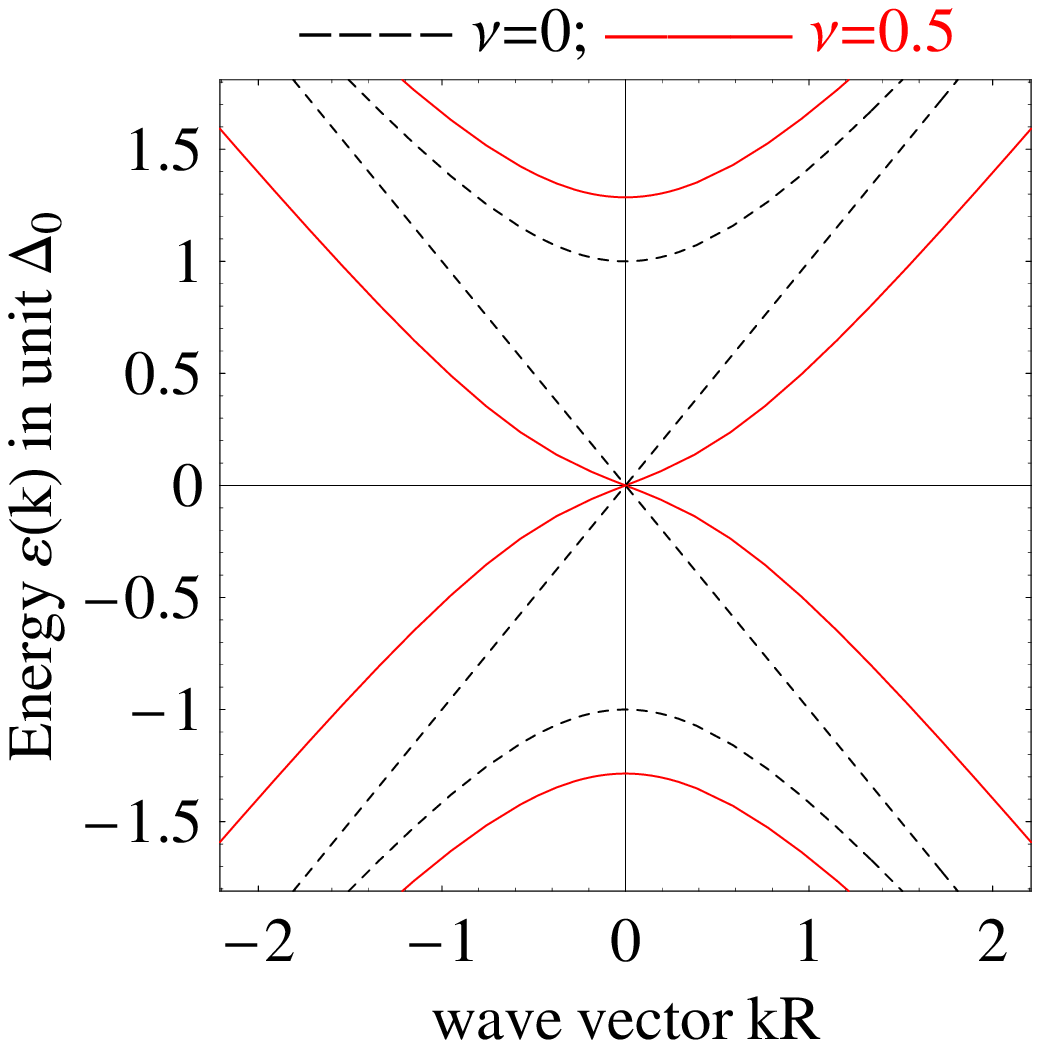}
\includegraphics*[width=0.70\linewidth]{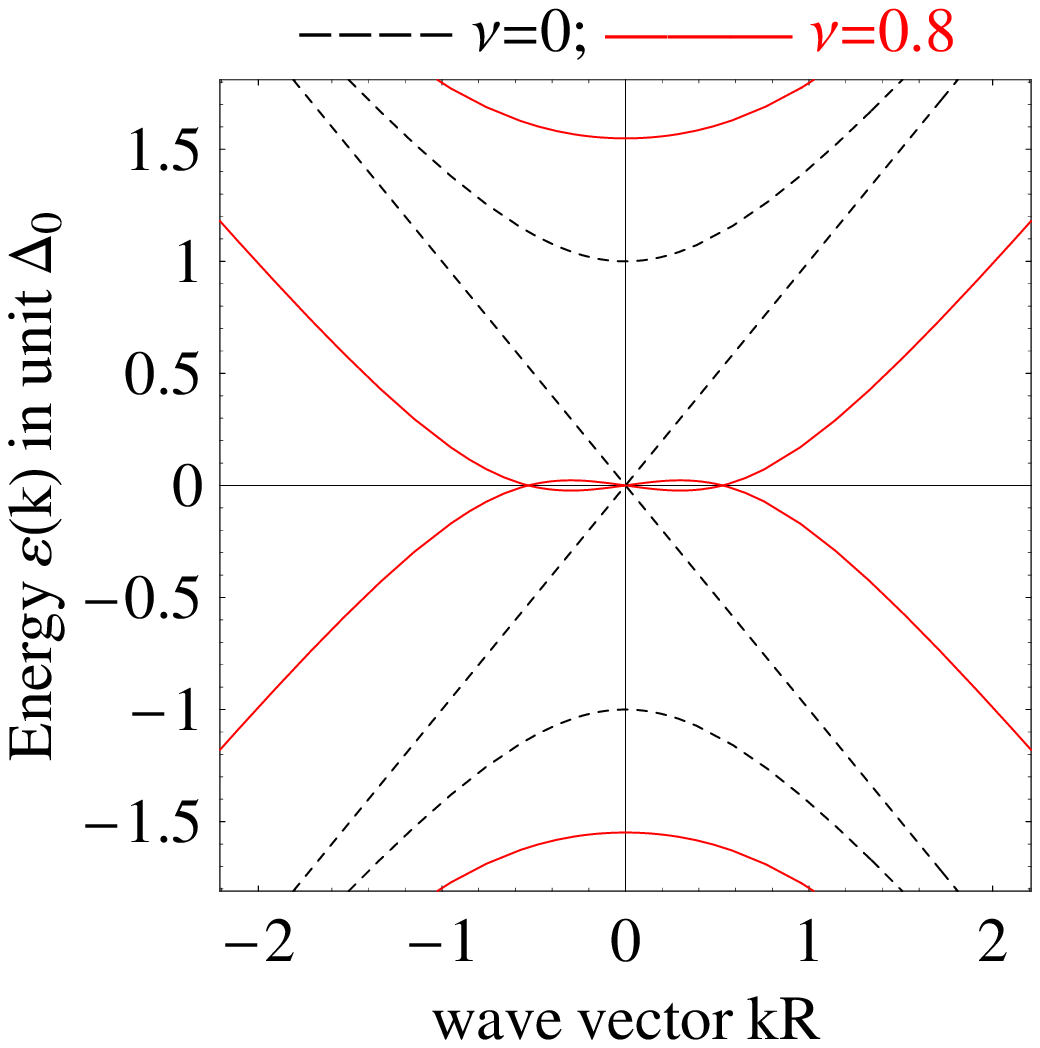}
 \caption{In the  x-axis the wavevector in unit $(k_y-K_s)R$ ($\pi_yR/\hbar$).
 (Top) Bandstructure of a non-metallic   CN with (red lines) and without (black dashed  lines)
 the transverse magnetic field ($\nu=0.5$).
 The main consequence of $B$  is the reduction of  the semiconducting
 gap.
 (Middle and bottom)
 Bandstructure of a metallic   CN with (red lines) and without (black dashed  lines)
 the transverse magnetic field.
 The main consequence of $B$  at intermediate fields is the rescaling
  of the Fermi velocity, while
 for quite strong fields a flat zone appears near $\pi_y=0$.
We know that  the magnetic parameter $\nu \approx 0.2$ for $B\sim
5\,T$ and $R\approx 50 nm$\cite{andos}.  }
\end{figure}
%%%%%%%%%%%%%%%%%%%%%%%%%%%%%%%%%%%%%%%%%%%%%%%%%%%%%%%%%%%%%

 From the expression of $
|{\Psi}_{m,\pm k}(\varphi,y)|^2$ we deduce a kind of "edge
localization" of the opposite current, analogous to the one
obtained for the QW\cite{noiqw} also for CNs.

Following the calculations reported in ref.(\cite{susy}) for a
metallic CN we can easily calculate the linear dispersion relation
changes near the band center $\epsilon=0$. Thus,
 the magnetic dependent  energy
can be written, near the Fermi points $k\sim K_s$, in terms of
$\nu$ as
 \be \label{g-metallic} \epsilon(|k-K_s|) = \pm \hbar |k-K_s| \left(
{v_F\over I_0(4\nu)} \right) \ . \ee
This describes a reduction of the Fermi velocity
$\hbar^{-1}d\epsilon/dk$ near $\epsilon=0$ by a factor
$I_0(4\nu)$.

Hence, the magnetic dependent Fermi wavevector follows
$$
k_F(\varepsilon_F,\nu,0) \approx K_s + \left(\frac{
\varepsilon_F}{\hbar v_F}\right)I_0(4 \nu),
$$
where the second term in the r.h.s.  depends on $B$ as
$$
k_F= K_s \pm k_0 + k(B)\approx K_s \pm k_0(1+4 \nu^2+...)
\rightarrow k(B) \sim   4 k_0 \nu^2,
$$
where $k_0= \left(\frac{ \varepsilon_F}{\hbar v_F}\right)$.

 \

{\it Electron-electron interaction - }

In order to analyze in detail the role of the e-e interaction, we
have to point out that  quasi 1D devices have low-energy branches,
at the Fermi level, that introduce a number of different
scattering channels, depending on the location of the electron
modes near the Fermi points. It has been often discussed that
processes which change the chirality of the modes, as well as
processes with large momentum-transfer (known as backscattering
and Umklapp processes), are largely subdominant, with respect to
those between currents of like chirality (known as forward
scattering processes)\cite{8n,9n,egepj}. This hierarchy of the
couplings characterizes the Luttinger regime. However in some
special cases the processes neglected here can be quite relevant,
giving rise even to a breakdown of the Luttinger
Liquid behaviour\cite{noisc}.

 \

Now, following Egger and Gogolin\cite{egepj}, we introduce the
unscreened Coulomb interaction in two dimensions
 \bea \label{U}
 U({\bf r}-{\bf r'})=\frac{c_0}{\sqrt{(y-y')^2+4
R^2 \sin^2(\frac{\varphi-\varphi'}{2})}}. \eea
  Then, we can
calculate $U_0(q,\omega_c)$  starting from  the eigenfunctions
$\widetilde{\Psi}_{0,k_F}(\varphi,y)$ and the potential in
eq.(\ref{U}). We focus our attention on the forward
scattering (FS) terms. We can obtain $g_2$, FS between opposite
branches, corresponding to the interaction between electrons with
opposite momenta, $\pm k_F$, with a small momentum transfer $\sim
q_c$. The strength of this term  reads \bea
 g_2&=&U_0(q_c,B,k_F,-k_F)\nonumber \\ &=&\frac{c_0}{ N_2(\nu)
}\left[ K_0(\frac{q_cR}{2})I_0(\frac{q_cR}{2}) +u_2(\nu)
K_1(\frac{q_cR}{2})I_1(\frac{q_cR}{2}) \right],\nonumber
\label{uq2} \eea where $K_n(q)$   denotes the modified Bessel
function of the second kind, $I_n(q)$ is  the modified Bessel
function of the first kind, while $N_2$ and $u_2$ are functions of
the transverse
  magnetic field, as we discuss in appendix. Analogously
  \bea
 g_4&=&U_0(q_c,B,k_F,k_F)\nonumber \\ &=&\frac{c_0}{ N_4(\nu)
}\left[ K_0(\frac{q_cR}{2})I_0(\frac{q_cR}{2}) +u_4(\nu)
K_1(\frac{q_cR}{2})I_1(\frac{q_cR}{2}) \right].\nonumber
\label{uq2} \eea

\section{Results}

{\it The bulk and the end critical exponents -}  The first result
of this paper concerns the dependence of the critical exponents on
the magnetic field, in large radius CNs.
   By introducing into  eq.(\ref{al1n})
the calculated values of $g_2$ and $g_4$, it follows that the bulk
critical exponent is reduced by the presence of a magnetic field,
as we show in Fig.(2).
%%%%%%%%%%%%%%%%%%%%%%%%%%%%%%%%%%%%%%%%%%%%%%%%%%%%%%%%%%%%%%%%%%%%%%%%%%%
\begin{figure}
\includegraphics*[width=1.0\linewidth]{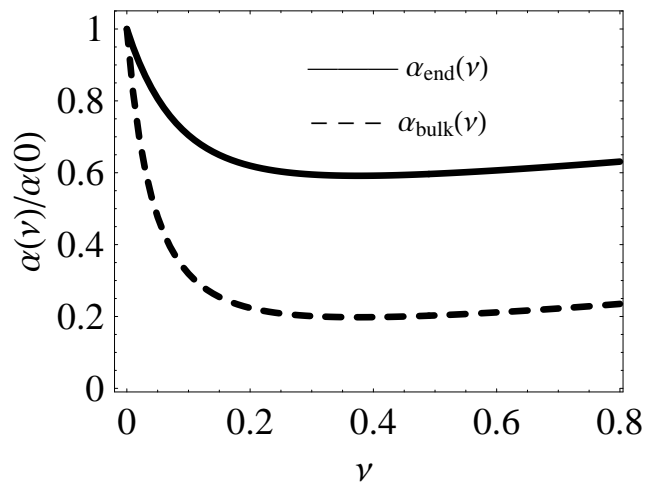}
\caption{{Critical exponents versus the magnetic field dependent
parameter, $\nu$, for a large radius  CN: $\alpha_{bulk}$ is
calculated following eq.(\ref{al1n}),
$\alpha_{end}$ is calculated following eq.(\ref{al2n}).% The magnetic field yields  a strong reduction of both the critical exponents.
%Now we are ready to discuss the effects of a magnetic field  that
%we can synthesize in few points.
%
%-
 The magnetic field rescales the values of the Fermi velocity and
the strength of e-e interaction. It follows  that the effects
of a transverse magnetic field also involve the value of $K$.
Thus, we predict a reduction of the critical exponents
$\alpha_{bulk}$ and  $\alpha_{end}$, by giving  magnetic field
dependent exponents for the power law behaviour of the
conductance.
 }}
\end{figure}
%%%%%%%%%%%%%%%%%%%%%%%%%%%%%%%%%%%%%%%%%%%%%%%%%%%%%%%%%%%%%%%%%%%%%%%%%%%%
%%%%%%%%%%%%%%%%%%%%%%%%%%%%%%%%%%%%%%%%%%%%%%%%%%%%%%%%%%%%%%%%%

This  prediction can be extended to $\alpha_{end}$, calculated
following eq.(\ref{al2n}), as we show in Fig.(2). Hence, it
follows that the exponent $\gamma_e-1$ can cross from positive to
negative values, when the magnetic field increases.

\

The experimental data about SWNT\cite{Postma01} gives, for
vanishing magnetic field,  $K_1\approx 0.26$,
$\alpha_{Bulk}\approx 0.27$ and $\alpha_{end}\approx 0.72$.

For a  MWNT we consider $N_s\sim 5$\cite{kanda,noiprl} so that
$K_5\approx 0.1$, $\alpha_{Bulk}\approx 0.2$ and
$\alpha_{end}\approx 0.4-0.5$.

\

{\it The intrinsic Quantum Dot  -} When there are some obstacles
to the free path of the electrons along a 1D device, a scattering
potential has to be introduced in the theoretical model.  The
presence of two barriers along a CN\cite{Postma01} at a distance
$d$  can be represented by a potential
$$V_B(y)=U_B\left(f(y+\frac{d}{2})+f(y-\frac{d}{2})\right),$$ where
$f(y)$ is a square barrier function, a Dirac Delta function or any
other function localized near $y=0$. In general we can  analyze
the single particle transmission in the presence of a magnetic
field, $t(\varepsilon_F,B)$,  by identifying the off-resonance
condition ($|t|=0$), where electrons are strongly backscattered by
the barriers, and the on-resonance condition ($|t|=1$), where  the
scattering at low temperatures is negligible.

 \

Hence, as shown in Fig.(3), where we report the transmission
$T=|t|^2$ versus $\nu$ for the lowest subband,  a magnetic field
dependent transmission follows, thus a magnetic dependence of the
peaks in the transmission is shown which exhibits
 a magnetically tuned transport through the CN. In particular, assuming
that there are two identical, weakly scattering barrier at a
distance $d$, the transmission is non-zero for particular values of
$k_F$, so that $\cos(k_F d)\approx 0$.
 %%%%%%%%%%%%%%%%%%%%%%%%%%%%%%%%%%%%%%%%%%%%%%%%%%%%%%%%%%%%%%%%%%%%%%%%%%%
\begin{figure}
\includegraphics*[width=1.0\linewidth]{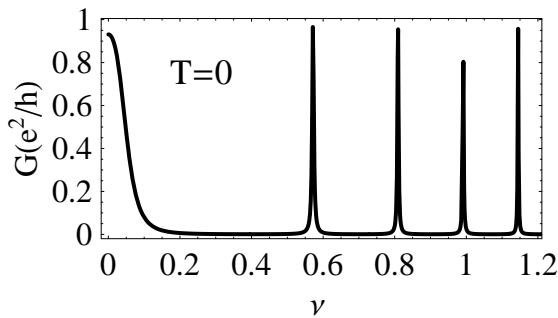}
\caption{Transmission ($T$) of the lowest subband ($m=0$)  as
a function of the magnetic field. We use a double square
barrier model, for the IQD. We  observe   the appearing of
resonance peaks, as a function of the magnetic field. The
ballistic conductance (at very low temperatures) is proportional to
$|t|^2$ according to the Landauer formula. }
\end{figure}
%%%%%%%%%%%%%%%%%%%%%%%%%%%%%%%%%%%%%%%%%%%%%%%%%%%%%%%%%%%%%%%%%%%%%%%%%%%%
%%%%%%%%%%%%%%%%%%%%%%%%%%%%%%%%%%%%%%%%%%%%%%%%%%%%%%%%%%%%%%%%%

We consider an intrinsic QD with $d\approx 250 nm$ in a CN of
$R\approx 5 nm$. Thus, starting from the electrons in the middle
of the bandgap, i.e. $\varepsilon_F\sim v_F \hbar/(2R)$,
$k(B)\approx \nu^2/(2R)$, we have to observe about $4$ peaks (i.e. the
number of resonances with $\nu \leq 1$ is $n_p=d/(4 \pi R)$) in
the transmission, when growing the magnetic field from $\nu=0$ to
$\nu=1$.

The presence of these oscillations has to  be seen in MWNTs or
SWNTs of large radius,  while in   the case of a  SWNTs with radius
$R\approx 1 nm$ the values of the magnetic field are  unrealistic.

\

Analogously to our previous paper\cite{noiqw} we can discuss the
different explanations of the resonance conditions. From a
theoretical point of view the on resonance condition can be seen
in two different ways: in some papers\cite{vw}, where
the ballistic transport in QWs was analyzed, it was discussed the presence of
these peaks as providing evidence of an Aharonov Bohm effect,
while in the TLL theory the resonance peaks are put in
correspondence to the presence of an average particle number
between the two barriers of the form $\nu+1/2$, with integer
$\nu$: thus we suppose that each electron in the QD carries a
quantum of magnetic flux.

\

{\it Temperature behaviour }- As it is known the presence of
the  peaks in the transmission has to be observable not only at
very low temperatures.  The temperature does not affect the values
of $B$ corresponding to  the conductance peaks, while their
largest value, $G_{max}$, follows a power law according to the
Sequential Tunneling theory. Thus, $G_{max}\propto T^{\gamma_e-1}$
with $\gamma_e$  depending on the tunneling mechanism. This point
deserves a brief discussion.

In this paper we take into account
a short nanotube section that is created by inducing (e.g. with
an atomic force microscope) local barriers into a large radius CN.
In this case the condition, $\Delta_{dot}\gg K_B T$ discussed in
ref.\cite{Postma01}  is confirmed in a large range of temperatures
around $T_R$ ($\Delta_{dot}/K_B\sim 10^4\; ^oK$, while
$\varepsilon_c/K_B\sim 1\; ^oK$).

Now we could discuss the two cases, by assuming the validity of
either the UST or the CST. In any case, we want to point out that
in both theories, it appears the critical exponent $\alpha_{end}$,
which has to be rescaled with the growing of the magnetic field.
The discussed reduction of $\alpha_{end}$, due to the increasing
magnetic field, also affects the shape of the peaks.

\

{\it The intersubbands processes -} The role of the many subbands
($N_s$) which cross the Fermi level should be taken into account
by introducing the matrix  $t_{n,m}$
 including all the intersubband scattering processes. However we can suppose  $|t_{n,m}|\ll
 |t_{n,n}|$, corresponding to the adiabatic regime, because the intersubbands
processes, i.e. the processes that involves two different
subbands, are largely subdominants with respect to the processes
involving the same subband. It follows that  the conductance $G$
results proportional to the sum of the $|t_{n,n}|$. Thus, the
peaks corresponding to the on resonance condition, due to the
$N_s$ subbands have to be superposed in order to calculate  the
zero temperature conductance. However, the contribution to the
oscillations due to the subbands different from the lowest one can
be  negligible, because the shift in $k(B)$ is quite smaller for
the higher subbands, as we show in Fig(1).

\

\section{Conclusions }

In this paper we extended to large radius CNs the formalism
introduced for a QW in a previous paper. We showed how the
presence of a magnetic field modifies the role played by both the
e-e interaction and the presence of obstacles in CNs of large
radius.

The first   prediction that comes from our study is that there
should be a significant reduction of the critical exponents, as
the magnetic field is increased, in agreement with the results
found for QWs.

Our second prediction concerns the presence of some peaks in the
small bias conductance versus the magnetic field.

It would be of considerable importance to test this behaviour  in
experiments carried out using different samples, in various
temperature regimes. This experimental test can be also useful, in
order to solve the controversial question about the exponent that
characterizes the power law dependence of $G(T)$.

We want to
remark that our approach is based on the idea that electrons
tunnel coherently through an obstacle, represented by a double
barrier, that can be assumed only as a strong barrier.

Our results could be surely affected by  the use of a model, where
the electrons weakly interact with the lattice, while the buckles
are represented by strong potential  barriers. This approximation
holds in the opposite regime, with respect to the model of
spinless fermions on the lattice used in ref.\cite{meden}.
However, we believe that our model can well reproduce some
experimental results
while, for what concerns the different regimes, we want also to
suggest that, when the temperature decreases, different approaches
could be needed, as we discussed in some of our previous
papers\cite{noicb,noiprl}.

\appendix
\begin{widetext}
\section{From the 2D Coulomb potential to a 1D Model}
Firstly, we introduce the wavefunctions $\Psi$  for a metallic CN,  as
spinors constructed starting from the functions
$$\widetilde{\psi}_{s,m, k}(\varphi,y)=\frac{e^{i\left(k y+ i m
\varphi\right)}}{%\sqrt{2\pi^2 L (2 \alpha_s^2 + \beta_s^2 +
%\gamma_s^2)}
N} \left(\alpha_s+ \beta_s \sin(\varphi)+\gamma_s \cos(\varphi))
\right),
$$
it follows
$$
\Psi^\dag \Psi=\frac{\sum_{s=\uparrow,\downarrow}\left((\alpha_s^2
+ \beta_s^2)+ 2 (\alpha_s\gamma_s) \cos(\varphi) +(\gamma_s^2 -
\beta_s^2) \cos(\varphi)^2 +\left[ 2 (\alpha_s\beta_s)
\sin(\varphi)+ (\beta_s\gamma_s) \sin(2 \varphi)\right] )
\right)}{{2\pi^2 L \sum_{s=\uparrow,\downarrow}(2 \alpha_s^2 +
\beta_s^2 + \gamma_s^2)}},
$$
and we define $\eta_\pm=\sum_{s=\uparrow,\downarrow} (\alpha_s^2 +
\beta_s^2)$, $\theta_\pm=\sum_{s=\uparrow,\downarrow}2
(\alpha_s\gamma_s)$, $
\xi_\pm=\sum_{s=\uparrow,\downarrow}(\gamma_s^2 - \beta_s^2)$ and
$N_\pm=\sum_{s=\uparrow,\downarrow}(2 \alpha_s^2 + \beta_s^2 +
\gamma_s^2)$, where $\pm$ corresponds to the values of $k=\pm
k_F$.

 Now we introduce the Coulomb interaction and expand
this function in terms of  $R/|y-y'|$ as
$$
U({\bf r}-{\bf r'})=\frac{c_0}{|y-y'|}\left(\sum_k^\infty
\frac{{\left( -1 \right) }^k\,\,\Gamma(\frac{1}{2} + k)}
  {{\sqrt{\pi }}\,\Gamma(1 + k)} \left(\frac{2R}{y-y'}\right)^{2\,k}\sin^{2\,k}(\frac{\varphi-\varphi'}{2})\right).
$$

The Forward scattering between opposite branches ($\pm$) is
obtained as \bea \nonumber
 U(y&-&y')\approx c_0 \int_{-\pi}^\pi
d\varphi\int_{-\pi}^\pi d\varphi'\sqrt{\frac{1}{(y-y')^2+4 R^2
\sin^2(\frac{\varphi-\varphi'}{2})}}{\Psi^\dag}_{m,k_F}(\varphi,y){\Psi}_{m,k_F}(\varphi,y){\Psi}_{n,-k_F}(\varphi',y'){\Psi^\dag}_{n,-k_F}(\varphi',y')
\\&=&\frac{c_0}{4\pi^2}\int_{-\pi}^\pi
d\varphi\int_{-\pi}^\pi d\varphi'\sqrt{\frac{1}{(y-y')^2+4 R^2
\sin^2(\frac{\varphi-\varphi'}{2})}} \left(\frac{\eta_++ \theta_+
\cos(\varphi' )+\xi_+\cos^2(\varphi'
)}{{N_+}}\right)\left(\frac{\eta_-+ \theta_- \cos(\varphi
)+\xi_-\cos^2(\varphi )}{{N_-}}\right)
\nonumber \\
       &\approx&\frac{c_0}{4\pi^2}\sqrt{\frac{1}{(y-y')^2}}\int_{-\pi}^\pi
d\varphi\int_{-\pi}^\pi d\varphi'\left(\sum_k^\infty \frac{{\left(
-1 \right) }^k\,\,\Gamma(\frac{1}{2} + k)}
  {{\sqrt{\pi }}\,\Gamma(1 + k)} \left(\frac{2R}{y-y'}\right)^{2\,k}\sin^{2\,k}(\frac{\varphi-\varphi'}{2})\right)
\nonumber
  \\ &\times&
  \left(\frac{(\eta_+ \eta_-+( \eta_+\theta_-+\eta_-\theta_+)\,\cos(\varphi )\,\cos(\varphi' ) +(\xi_+\cos^2(\varphi )+\xi_-\cos^2(\varphi'
  )))}{{\left(N_+N_-
        \right) }}\right)
\nonumber \\
        & =&
        \frac{2 c_0}{N_+N_-}\sqrt{\frac{1}{(y-y')^2}}\sum_n^\infty \frac{{\left(
-1 \right) }^n\,\,\Gamma(\frac{1}{2} + n)}
  {{\sqrt{\pi }}\,\Gamma(1 + n)} \left(\frac{2R}{y-y'}\right)^{2\,n}\nonumber \\
  &\times& \left( \left(\eta_+ \eta_+ + \frac{\xi_++\xi_-}{2}\right)\frac{4\pi^{3/2}\Gamma(n+1/2)}{\Gamma(n+1)}+( \eta_+\theta_-+\eta_-\theta_+)\frac{2\pi^{3/2}n \Gamma(n+1/2)}{\Gamma(n+2)}
  \right).
 \eea
 Thus, we introduce $u_0=\eta_+ \eta_-$,$u_1=(
 \eta_+\theta_-+\eta_-\theta_+)$and $u_2=(\xi_++\xi_-)$
so that we obtain
 \bea
 U(y-y')&=& 2 \frac{c_0}{\left( N_+N_-\right)}\sqrt{\frac{1}{(y-y')^2}}\nonumber \\
        &\times& \left\{ (u_0+\frac{u_1}{2}) K(-(\frac{2R}{y-y'})^2) + u_2
 \left( \frac{\pi}{8} \, _2F_1(\frac{3}{2},\frac{3}{2};2,-(\frac{2R}{y-y'})^2)\left[\frac{2R}{y-y'}\right]^2\right)\right\}
 \eea
 Where  $K_E(x)$ gives the complete elliptic integral of the first
kind while $_2F_1(a,b,c,z)$  is the hypergeometric function.

The Fourier Transform gives the $U_0(q)$ as \bea
U_0(q)=\frac{c_0}{\sqrt{2}\left( N_+ N_-
\right)^2}\left[(u_0+\frac{u_1}{2})
K_0(\frac{qR}{2})I_0(\frac{qR}{2}) +\frac{u_2}{2}
K_1(\frac{qR}{2})I_1(\frac{qR}{2}) \right] \eea with $K_n(q)$
which gives the modified Bessel function of the second kind and
$I_n(q)$ gives the modified Bessel function of the first
kind.

In order to calculate $g_4$ we have to define  $u_0=\eta_+^2
$,$u_1=2 \eta_+\theta+$ and $u_2=2\xi_+$, and then plug these expressions in the
equations above.
\end{widetext}

%%%--------------------------------------------------------

%%%------------------------------------------------------------------------
\bibliographystyle{prsty} %Phys. Rev. style
\bibliography{}

\begin{thebibliography}{99}
%%%%%%%%%%%%%%%%%%-
\expandafter\ifx\csname
natexlab\endcsname\relax\def\natexlab#1{#1}\fi
\expandafter\ifx\csname bibnamefont\endcsname\relax
\def\bibnamefont#1{#1}\fi
\expandafter\ifx\csname bibfnamefont\endcsname\relax
\def\bibfnamefont#1{#1}\fi
\expandafter\ifx\csname citenamefont\endcsname\relax
\def\citenamefont#1{#1}\fi
\expandafter\ifx\csname url\endcsname\relax
\def\url#1{\texttt{#1}}\fi
\expandafter\ifx\csname
urlprefix\endcsname\relax\def\urlprefix{URL }\fi
\providecommand{\bibinfo}[2]{#2}

\bibitem{noiqw}
 S.
Bellucci and P. Onorato, {Eur. Phys. J. B} {\bf 47}, 385-390 (2005).


\bibitem{TL} S.~Tomonaga, Prog. Theor. Phys. {\bf 5}, 544 (1950);
J.~M.~Luttinger, J.~Math. Phys. {\bf 4}, 1154 (1963); D.~C.~Mattis
and E.~H.~Lieb, J.~Math. Phys. {\bf 6}, 304 (1965).
\bibitem{wires}
A. Yacoby, H. L. Stormer, N. S. Wingreen, L. N. Pfeiffer, K. W.
Baldwin and K. W. West, Phys.\ Rev.\ Lett.\ {\bf 77}, 4612 (1996);
O. M. Auslaender, A. Yacoby, R. de Picciotto, K. W. Baldwin, L. N.
Pfeiffer and K. W. West, Phys.\ Rev.\ Lett.\ {\bf 84}, 1764
(2000),%; M. Rother, W. Wegscheider, R. A. Deutschmann, M. Bichler
%and G. Abstreiter, Physica~E {\bf 6}, 551 (2000).

\bibitem{cnts}
S. J. Tans, M. H. Devoret, H. Dai, A. Thess, R. E. Smalley, L. J.
Geerligs and C. Dekker, Nature {\bf 386}, 474 (1997);% M. Bockrath,
%D. H. Cobden, J. Lu, A. G. Rinzler, G. Andrew, R. E. Smalley, L.
%Balents and P. L. McEuen, Nature {\bf 397}, 598 (1999);
Z. Yao, H.
W. J. Postma, L. Balents, and C. Dekker, Nature {\bf 402}, 273
(1999).

\bibitem{kf}
C. L. Kane and M. P. A. Fisher, Phys. Rev. Lett. {\bf 68}, 1220
(1992); C. L. Kane and M. P. A. Fisher, Phys. Rev. B {\bf 46},
R7268 (1992)

\bibitem{sh}
H. J. Schulz, cond-mat/9503150.

\bibitem{Postma01} H.W.Ch. Postma, T. Teepen, Z. Yao, M. Grifoni, and C. Dekker,
Science {\bf 293}, 76 (2001).

\bibitem{Bozovic01} D. Bozovic, M. Bockrath, J.H. Hafner, C. M. Lieber, H. Park,
and M. Tinkham, Appl. Phys. Lett. {\bf 78}, 3693 (2001).


\bibitem{Sassetti95} M. Sassetti, F. Napoli, and U. Weiss, Phys. Rev. B
{\bf 52}, 11213 (1995).
\bibitem{Furusaki98} A. Furusaki, Phys. Rev. B {\bf 57}, 7141 (1998).

\bibitem{Braggio00} A. Braggio, M. Grifoni, M. Sassetti, and F. Napoli,
Europhys. Lett. {\bf 50}, 236 (2000).

\bibitem{Thorwart02} M. Thorwart, M. Grifoni, G. Cuniberti, H.W.Ch. Postma,
and C. Dekker, Phys. Rev. Lett. {\bf 89}, 196402 (2002).

\bibitem{Nazarov03} Yu.V. Nazarov and L.I. Glazman,
Phys.  Rev. Lett. {\bf 91}, 126804 (2003).

\bibitem{Polyakov03} D.G. Polyakov and I.V. Gornyi,
Phys. Rev. B {\bf 68}, 035421 (2003).

\bibitem{Komnik03} A. Komnik and A.O. Gogolin,
Phys. Rev. Lett. {\bf 90}, 246403 (2003).

\bibitem{Hugle04} S. H\"ugle and R. Egger, Europhys. Lett. {\bf 66}, 565 (2004).

\bibitem{furusaki_double_barriere}
A. Furusaki and N. Nagaosa, Phys. Rev. B {\bf 47},  3827  (1993).
\bibitem{meden}
V. Meden,T. Enss,  S. Andergassen, W. Metzner, and K. Sch\"onhammer,  Phys. Rev. B {\bf 71}, 041302(R) (2005),
\bibitem{3n}
 J.W. Mintmire, B.I. Dunlap, C.T. White, Phys. Rev. Lett. {\bf 68} (1992) 631;
N. Hamada, S. Sawada, A. Oshiyama, Phys. Rev. Lett. {\bf 68} (1992)
1579; R. Saito, M. Fujita, G. Dresselhaus, M.S. Dresselhaus, Appl.
Phys. Lett. {\bf 60} (1992) 2204.

\bibitem{[21]}
M. Kr\"uger, M. R. Buitelaar, T. Nussbaumer, C. Sch\"onenberger,
and L. Forr\'o, {\em Appl. Phys. Lett.} {\bf 78}, 1291 (2001).

\bibitem{npb}
S. Bellucci, J. Gonz\'alez, P. Onorato, {Nucl. Phys. B} {\bf
663} [FS] (2003) 605; S. Bellucci, J. Gonz\'alez, P. Onorato, {
Phys. Rev. B} {\bf 69} (2004) 085404.
\bibitem{noiqw0}
 S. Bellucci and P. Onorato, {Eur. Phys. J. B} {\bf 45}, 87-96
(2005).


\bibitem{egmw} R. Egger, Phys.\ Rev.\ Lett.
{\bf 83}, 5547 (1999).

\bibitem{B}
A. Bachtold, M. de Jonge, K. Grove-Rasmussen, P. L. McEuen, M.
Buitelaar and C. Schænenberger, {\em Phys. Rev. Lett.}  {\bf 87},
166801 (2001);\\
A. Bachtold, M. de Jonge, K. Grove-Rasmussen, P.L. McEuen, M.
Buitelaar and C. Schonenberger, report cond-mat/0012262.


\bibitem{y}
Z. Yao, H. W. Ch. Postma, L. Balents and C. Dekker, {\em Nature}
{\bf 402}, 273 (1999).
\bibitem{[33]}
H. Ajiki and T. Ando, J. Phys. Soc. Jpn. 62, 1255 (1993).

\bibitem{[34]}
A. Kanda et al., Physica B, {\bf 323}, 107-114   (2002).
%K. Tsukagoshi, N. Yoneya, S. Uryu, Y. Aoyagi, A. Kanda, Y. Ootuka, et al., Physica B 323 (2002).

\bibitem{kanda}
A. Kanda, K. Tsukagoshi, Y. Aoyagi and Y. Ootuka, \prl {\bf 92},
36801 (2004).
%\bibitem{noiqw}
% S. Bellucci and P. Onorato, {Eur. Phys. J. B} {\bf 45}, 87-96
%(2005).

\bibitem{noimf}
S. Bellucci and P. Onorato, Annals of Physics  {\bf 321}, 934-949
(2006).
\bibitem{susy}
H.-W. Lee, D. S. Novikov, Phys. Rev. B {\bf 68} 155402 (2003).

\bibitem{andos}
T. Ando, T. Seri, J. Phys. Soc. Jpn. {\bf 66}, 3558 (1997).

\bibitem{8n}
L. Balents, M.P.A. Fisher, Phys. Rev. B {\bf 55} (1997) R11973.
\bibitem{9n}
R. Egger, A. O. Gogolin, Phys.\ Rev.\ Lett.\ {\bf 79} (1997) 5082
\bibitem{egepj}
R. Egger, A. O. Gogolin,, Eur. Phys. J. B {\bf 3} (1998) 281.
\bibitem{noisc}
S. Bellucci, M. Cini ,P. Onorato and E. Perfetto  (2006) accepted
for publication in Journal of Physics: Condensed Matter
cond-mat/0603853.
\bibitem{vw}
B. J. van Wees, H. van Houten, C. W. J. Beenakker, J. G.
Williamson, L. P. Kouwenhoven, D. van der Marel and C. T. Foxon,,
Phys. Rev. Lett. {\bf 60}, 848 (1988).

\bibitem{noicb}
S. Bellucci and P. Onorato, Phys. Rev. B {\bf 71}, 075418 (2005).
\bibitem{noiprl}
S. Bellucci, J. Gonz\'alez, P. Onorato, Phys. Rev. Lett. {\bf 95},
186403 (2005).

%%%%%%%%%%%%%%%%%%%%%%%%%%%%%%%%%%%%%%%%%%%%%%%%%%%%%%%%%%%%%%%%%%%%%%%%%%%%%%%%%%%%%%

%%%%%%%%%%%%%%%%%%%%%%%%%%%%%%%%%%%%%%%%%%%%%%%%%%%%%%%%%%%%%%%%%%%%%%%%%%%%%%%%%%%%%%

%%%%%%%%%%%%%%%%%%%%%%%%%%%%%%%%%%%%%%%%%%%%%%%%%%%%%%%%%%%%%%%%%%%%%%%%%%%%%%%%%%%%%%


\end{thebibliography}

\end{document}